\documentstyle[12pt]{article}
\begin{document}

\def\beq{\begin{equation}}
\def\eeq{\end{equation}}  
\def\bea{\begin{eqnarray}}
\def\eea{\end{eqnarray}}  
\def\bq{\begin{quote}}    
\def\eq{\end{quote}}      
\def\ra{\rightarrow}      
\def\lra{\leftrightarrow} 
\def\ups{\upsilon}
\def\nn{\nonumber}
\def\r.{\right.}
\def\l.{\left.}
\def\o.{\overline}
\newcommand{\sheptitle}
{Infra-Red Stable Fixed Points of Yukawa Couplings in Supersymmetric Models}
\newcommand{\shepauthor}
{B. C. Allanach$^1$ and
S. F.
King$^2$ \\ 
\vspace{\baselineskip}
{\small 1. Rutherford Appleton Laboratory, Chilton, Didcot, OX11 0QX, U.K.} \\
{\small 2. Department of Physics and Astronomy, University of
Southampton, Southampton, SO17 1BJ, U.K.}}

\newcommand{\shepabstract}
{We provide the solutions of the fixed point conditions of the Yukawa sector
for a large class of $N=1$ supersymmetric theories including the
minimal and next-to-minimal supersymmetric standard models
and their grand unified and other extensions. 
We also introduce a test which can discriminate between infra-red stable,
infra-red unstable and saddle point solutions, and illustrate our methods
with the example of the next-to-minimal supersymmetric
standard model. We show that in this case, the fixed point prediction of the
top quark mass is equivalent to that of the minimal supersymmetric standard
model, supporting previous numerical analyses.}

\begin{titlepage}
\begin{flushright}
RAL TR-97-019 \\ SHEP 97-07 \\ 
\end{flushright}
\vspace{.4in}
\begin{center}
{\large{\bf \sheptitle}}
\bigskip \\ \shepauthor \\ \mbox{} \\
\vspace{.5in}
{\bf Abstract} \bigskip \end{center} \setcounter{page}{0}
\shepabstract
\end{titlepage}

The fermion mass problem in the standard model arises from the
presence of many unknown dimensionless Yukawa couplings.
A possible solution to this long-standing problem is that the 
standard model could be embedded in some more predictive high energy
theory with fewer Yukawa couplings. Alternatively one may attempt to
relate the Yukawa couplings to the gauge coupling, 
as in the Pendleton-Ross infra-red stable 
fixed point (IRSFP) for the top quark Yukawa coupling \cite{RPtop},
or the quasi-fixed point of Hill \cite{SMtop}\footnote{The relation between
the fixed point and 
quasi-fixed point has been fully elucidated by Lanzagorta and Ross
\cite{GG}.}.
To obtain IRSFPs it is necessary that the unknown
dimensionless Yukawa couplings are of the same order 
as the gauge coupling at high energy. Thus at first sight this
approach would seem to be inapplicable to the small Yukawa couplings
of the standard model.
However string theory yields large Yukawa couplings of order the
gauge coupling, and this
could suggest that the small dimensionless
Yukawa couplings be reinterpreted as large Yukawa couplings 
multiplied by a ratio of mass scales. In such a scenario
fixed points may once again be applicable, as pointed out recently by
Ross \cite{GGfixed}. In this case the high energy scale would be the
string scale and the infra-red region would be near the GUT scale.
Thus IRSFPs may provide an unexpected resolution of the
fermion mass problem. IRSFPs in the MSSM can
also be used to explain the apparent success of the $\lambda_\tau=\lambda_b$
GUT scale relation of some popular SUSY GUTs~\cite{schrempp}.
Further fixed points have also been identified in the soft masses of
supersymmetric theories~\cite{Rosssoft,Jones,symbreak}. It is clear that the
idea 
of IRSFPs has general applicability, and here we shall follow a 
model independent approach.

In this letter 
we shall present a general approach which may be used to
identify the fixed points that exist in the dimensionless
Yukawa couplings of a large class of
$N=1$ supersymmetric theories, and to determine the stability properties
of the fixed points in the infra-red.
Our assumptions are listed below:
\begin{enumerate}
\item{We work at one loop order only.}
\item{We assume that the wavefunction anomalous dimensions are
diagonal $\gamma_i^j\propto \delta_i^j\gamma_i$. This may be due to
some symmetry that forbids wavefunction mixing between
different irreducible representations. For example
in the MSSM R-parity forbids mixing between the lepton doublet $L$ and
the Higgs doublet $H_1$.}
\item{We assume a single gauge coupling. This may either be due to
several approximately equal gauge couplings, 
which could be the case near the string scale, or due to there being a
single dominant
gauge coupling as in the case of QCD at low energies where the electroweak
couplings may be neglected.}
\end{enumerate}

Our general approach may be compared to that of 
Lanzagorta and Ross \cite{GG} for a single Yukawa coupling, 
however unlike these authors
we shall present a well defined procedure for obtaining the
fixed points for theories involving
any number of large Yukawa couplings in the theory,
and having identified the fixed points we shall give a general discussion
of their infra-red stability.

We shall be interested in the
one-loop renormalisation group equations (RGEs) for the dimensionless
trilinear Yukawa couplings of an N=1 SUSY theory,
which are by now well known and have been widely reported
(see for example Martin and Vaughn~\cite{MandV}).
The trilinear part of the superpotential may be written
\beq
W = \frac{1}{6} h^{ijk} \phi_i \phi_j \phi_k,
\eeq
where $h^{ijk}$ are the dimensionless Yukawa couplings, $\phi_i$ represent the
superfields and the sums over $i,j,k$ run over all superfields in the spectrum
of the effective theory. The relevant RGEs read
\bea
16 \pi^2 \frac{d h^{ijk}}{d \ln \mu} &=& h^{ijp} \gamma_p^k + h^{ipk} \gamma_p^j +
h^{pjk} \gamma_p^i, \label{MVRGE} \\
16 \pi^2 \frac{d g_a}{d \ln \mu} &=& b_a g_a^3 \label{Bfns}
\eea
where $\mu$ is the $\overline{MS}$ renormalisation scale and
the anomalous dimension is defined as
\beq
\gamma_i^j \equiv \frac{1}{2} \sum_{p,q}h_{ipq} h^{jpq} 
- 2 \delta^j_i g_a^2 C^a_i
\label{anom} 
\eeq
to one loop order. $b_a$ is the one-loop beta function of the gauge group
$G_a$ and
the sum over $a$ here represents a sum over all dominant
gauge couplings. $C^a(i)$ is the quadratic Casimir of the representation of
$\phi_i$ under the group with gauge coupling $g_a$:
\beq
(T^a_R T^a_R)^j_i \equiv C^a_R \delta^j_i.
\eeq
Here, $T^a_R$ is a matrix in the $R$
representation for the group labeled by $a$.
Pictorially, the RG evolution of the trilinear coupling can be described as
an insertion of the anomalous dimension correction on each external
leg~\cite{us}. 

According to our second assumption, the wavefunction anomalous dimensions
are diagonal,
\beq
\gamma_i^j=\gamma_i \delta_i^j
\eeq
then the above equations become:
\bea
16 \pi^2 \frac{d h^{ijk}}{d \ln \mu } &=& h^{ijk}( \gamma_i + \gamma_j +
\gamma_k ), \label{MVRGE1}
\eea
where the anomalous dimension is defined as
\beq
\gamma_i  \equiv \frac{1}{2} \sum_{p,q}h_{ipq} h^{ipq} - 2 g_a^2 C^a_i
\label{anom1} 
\eeq
We now introduce a more convenient notation:
\beq
Y^{ijk}\equiv \frac{(h^{ijk})^2}{16\pi^2}, 
\ \ \tilde{\alpha_a} \equiv \frac{g_a^2}{16\pi^2}, 
\ N_i \equiv -\frac{\gamma_i}{16\pi^2}, \ t \equiv -\ln \mu^2
\label{convenient}
\eeq
in terms of which:
\bea
\frac{d Y^{ijk}}{dt} &=& Y^{ijk}( N_i + N_j + N_k ). \label{MVRGE2}
\eea

According to our third assumption there is a single gauge coupling 
in the theory.
We define the ratio of each Yukawa coupling to the single
gauge coupling $\alpha$ as:
\beq
R^{ijk}=\frac{Y^{ijk}}{\tilde{\alpha}}
\label{ratio}
\eeq
The RGEs now become:
\bea
\frac{d R^{ijk}}{dt} &=& {\tilde{\alpha}}R^{ijk}
( N_i + N_j + N_k +{\tilde{\alpha}}b), \label{MVRGE3}
\eea
The non-trivial (non-zero) IRSFPs we seek correspond to 
$\frac{d R^{ijk}}{dt}=0$ for all $i,j,k$ with 
$R^{ijk}\neq 0$, which implies:
\beq
N_i + N_j + N_k +{\tilde{\alpha}}b =0, \label{IRSFP}
\eeq
for all $i,j,k$ corresponding to $R^{ijk}\neq 0$.

In the above notation the anomalous dimension
$N_i$ refers to the superfield $\phi_i$ 
while the trilinear coupling between three such superfields
is written as a three index tensor $h^{ijk}$. 
In a given theory each superfield will in general be in some representation
of the gauge group, and $i$ runs over all gauge indices as well as
the different representations themselves.
Thus most of the entries of the
tensor $h^{ijk}$ will be zero due to gauge invariance and other
imposed symmetries, and many of the non-zero entries will be equal
by gauge invariance. For example in a toy theory with a single 
top quark Yukawa coupling, assuming that QCD is the only
gauge group, the index $i$ would run over the 6 components
of the quark doublet $Q=(t,b)$, the 3 components of the top 
field $t^c$ and the 2 components of the Higgs doublet $H$, 
and in general the Yukawa coupling would be $h^{ijk}$ which 
would be a three index tensor with each index taking 11 possible values.
The top quark Yukawa coupling is characterised by 
a single parameter $h$, as is clear by dropping all indices
and writing the superpotential $W=hQt^cH_2$. In looking for
IRSFPs we are interested in the value of $h$, so it is clear that we
must simplify the notation in order to remove all the redundant
information contained in the tensor $h^{ijk}$ to arrive at the
physical coupling $h$. In the case of the top quark Yukawa coupling
it is clear that we can define $Y_h\equiv h^2/(16\pi^2)$ and
write the RGE for the single physical top quark Yukawa coupling as
\beq
\frac{dY_h}{dt} = Y_h(N_Q+N_{t^c}+N_{H_2})
\eeq
or defining $R_h\equiv Y_h/\tilde{\alpha}$,
\beq
\frac{dR_h}{dt} =  R_h(N_Q+N_{t^c}+N_{H_2}+{\tilde{\alpha}}b).
\eeq
where
\bea
N_Q     & = & 2C_Q\tilde{\alpha} -Y_h \nonumber \\
N_{t^c} & = & 2C_{t^c}\tilde{\alpha} -2Y_h \nonumber \\
N_{H_2} & = & -3Y_h
\eea
and the QCD Casimirs are $C_Q=C_{t^c}=4/3$.
The RGE in this case is thus:
\beq
\frac{dR_h}{dt} = {\tilde{\alpha}}R_h[(r_h+b)-6R_h ].
\eeq
where
\beq
r_h\equiv \sum_R 2C_R = 16/3.
\eeq
and in the MSSM the QCD beta function is $b=-3$, which leads to the
Pendleton-Ross fixed point $R_h^*=7/18$.

In general we can always relabel the Yukawa couplings of any theory as
$h_i$ where $i=1,\cdots n$ now runs over the $n$ non-zero physically
distinct couplings in the theory, and should not be confused with
the previous use of the index $i$.
Defining $Y_i\equiv h_i^2/(16\pi^2)$ and
$R_i\equiv Y_i/\tilde{\alpha}$, we
can always write the RG equations for the
physically distinct Yukawa couplings
of the theory as
\begin{equation}
\frac{d R_i}{d t} = \tilde{\alpha} R_i \left[ (r_i+b) - \sum_j S_{ij}
R_j
\right], \label{RGER}
\end{equation}
where $r_i = \sum_R 2C_R$ with the sum being
over the representations of the three fields associated with the
trilinear coupling $h_i$, and $S_{ij}$ is some
matrix whose numerical value is fully specified by
the wavefunction anomalous dimensions. 
The fixed point is then reached when the 
right hand side of Eq.\ref{RGER} is zero for all $i$. 
Thus, writing the fixed point solutions as $R_i^*$
there are two fixed point values for each coupling: $R_i^*=0$ or the
solution corresponding to
\beq
[ (r_i+b) - \sum_j S_{ij} R_j^*]=0
\eeq
The non-trivial fixed point solution is:
\beq
R_i^* = \sum_{j=1}^{n} ({S}^{-1})_{ij} (r_j+b) 
\label{nontrivial}
\eeq
We shall begin our discussion by considering only the 
non-trivial fixed points, and later extend our analysis to the
general case where some of the couplings are zero at the fixed point.

To determine the infra-red stability of the system in Eq.\ref{RGER}, we need
to Taylor expand it around the fixed point given in Eq.\ref{nontrivial}.   
We can then drop all except the linear terms, the resulting system of which 
allows an algebraic solution and can
thus be tested for infra-red stability.
We therefore make a change of variables to $\rho_i(t) \equiv R_i(t) - R_i^*$.
The RGE Eq.\ref{RGER} then becomes
\bea
\frac{d \rho_i(t)}{d t} &=&  \tilde{\alpha}(t) (\rho_i(t)+ R_i^*)
\left[ (r_i+b_i) - \sum_{j=1}^n{S}_{ij}(\rho_j(t)+R_j^*) \right]\label{Taylor}
\eea
where we have substituted the fixed point values of $R_i^*$ from
Eq.\ref{nontrivial}.

When we drop the quadratic terms in Eq.\ref{Taylor} and change the independent
variable from $t$ to $\tilde{\alpha}$ by Eq.\ref{Bfns}, we obtain the
linearised system
\beq
\frac{d \rho_i(t)}{d \ln \tilde{\alpha}(t)} \approx \frac{1}{b} R_i^*
\sum_{j=1}^n S_{ij}
\rho_j(t) \label{linearised}
\eeq
(no sum on $i$.)
Eq.\ref{linearised} then describes the behaviour of the trajectories as they
approach the fixed point. It has a general solution: 
\beq
\rho_i(t) \approx \sum_{k=1}^n a_k x^{(k)}_i 
\left( \tilde{\alpha}(t)\right)^{\lambda_k},
\label{solns2}
\eeq
where $x^{(k)}_j, \lambda_k$ are the $k=1,\cdots n$ eigenvectors
and eigenvalues of the eigenvalue equation
\begin{eqnarray}
\sum_{j=1}^n A_{ij} x^{(k)}_j &=& \lambda_k x^{(k)}_i \nonumber \\
A_{ij} &\equiv& \frac{1}{b} R_i^* S_{ij}. \label{evalue}
\end{eqnarray}
The $a_k$ in the solutions are constants 
which describe the particular linear combination of eigenvectors and
are set by initial boundary conditions.
The infra-red stability properties of the solutions in Eq.\ref{solns2} 
are independent of the
sign of the one-loop gauge beta function $b$, as we shall now show.

For $b>0$, $\tilde{\alpha}$ decreases with decreasing renormalisation scale
$\mu$. We require every eigenvalue
$\lambda_k$ of the matrix $\frac{1}{|b|} R_i^* S_{ij}$
to have a positive real part for infra-red stability, so that
$\rho_i \rightarrow 0$ as $\mu$ decreases.
For $b<0$, $\tilde{\alpha}$ increases with decreasing renormalisation scale
$\mu$, in this case we require all the eigenvalues of the matrix 
$-\frac{1}{|b|} R_i^* S_{ij}$ 
to have negative real parts, which is equivalent to the previous condition
for $b>0$, so the stability condition is independent of the
sign of the beta function, as claimed.
$\lambda_k=0$ corresponds to a direction in coupling space which is neither
attracted to, nor repelled by the fixed point (to first order).
For each of these directions there
should be one free parameter in the solution to the fixed point equations
which embodies the information of where a solution lies along this
direction (and is set by the initial boundary conditions). This corresponds to
some
information about the higher energy physics being retained at lower energies.

Complex eigenvalues always come in complex conjugate pairs, as do their
associated eigenvectors. Writing $\lambda_k \equiv t_k+is_k$, where $t_k,s_k$
are
real, the solution in this case is
\beq
\rho_i \approx\sum_{k} a_k
(x^{(k)}_i \tilde{\alpha}^{t_k + is_k} 
+ {x^{(k)}_i}^* \tilde{\alpha}^{t_k-is_k})
= \sum_{k}a_k
\tilde{\alpha}^{t_j} \left[ x^{(k)}_i \tilde{\alpha}^{i s_k} + {x^{(k)}_i}^* 
\tilde{\alpha}^{-is_k} \right] \label{spiral}
\eeq
where the sum runs over all values of $k$ corresponding to pairs of complex
eigenvalues. 
Eq.\ref{spiral} describes a spiral-like solution, the distance to the fixed
point being controlled by $\tilde{\alpha}^{t_k}$. Thus $t_k$ must be positive
for the trajectory to be infra-red stable.
For $b<0$, $\tilde{\alpha}$ increases with decreasing normalisation scale so
we
require $Re(\lambda_k) < 0$ for all $k$ for the solution to be
stable in the infra-red direction. 
If any of the conditions are not met, the fixed point is either a saddle
point or
an ultra-violet fixed point and so the fixed point will never be achieved at
low energies and must be rejected. In such cases we are led to consider
other fixed points where some of the couplings take
zero values at the fixed point. Another motivation for studying such 
fixed points would be if the non-trivial fixed point solution
led to values of $R_i$ with unphysical negative values
(see example later).

As already remarked, in general for each value of $i$ there are
two possible fixed point solutions to consider: $R_i^*=0$
or its non-zero fixed point value considered above.
There are thus $2^n$ fixed points contained in the theory (some of which could
be degenerate), and so
far we have only considered one possibility (the non-zero case
for each coupling). For each of the remaining possibilities it is
straightforward to determine the fixed points. Suppose in general that
we are considering a possibility with $m$ zero fixed point solutions
and $n-m$ non-zero solutions. First, we
re-order the $n$ couplings such that
\bea
\left[ R_i^* \right] _{i=1,\ldots,m } &=& 0 \nn \\
\left[ R_i^* \right] _{i=m+1,\ldots,n } &=& 
\sum_{j=m+1}^{n} ({S}^{-1})_{ij} (r_j+b) 
\label{ooaacantona}
\eea
Note that the non-zero couplings are now determined by
the lower right hand $(n-m)\times(n-m)$
block of the re-ordered matrix $S$. 
As we shall see the discussion of the 
non-zero solutions turns out to be 
similar to that previously 
considered, while that of the zero fixed points is even simpler.

For the general case including $m$ zero couplings, we need
to Taylor expand around the fixed point given in Eq.\ref{ooaacantona}.   
We make a change of variables to $\rho_i(t) \equiv R_i(t) - R_i^*$,
where $R_i^*=0$ for $i=1,\cdots m$.
The RGE Eq.\ref{RGER} then becomes
\bea
\left[ \frac{d \rho_i(t)}{d t} \right] _{i=1,\ldots ,m } & = &
\tilde{\alpha}(t) \rho_i(t)
\left[ (r_i+b) 
- \sum_{j=1}^m S_{ij}\rho_j(t) 
- \sum_{j=m+1}^n {S}_{ij}(\rho_j(t)+R_j^*) 
\right]
\nn \\
\left[ \frac{d \rho_i(t)}{d t} \right] _{i=m+1,\ldots,n } & = &   
\tilde{\alpha}(t) (\rho_i(t)+ R_i^*)
\left[ (r_i+b) 
- \sum_{j=1}^m S_{ij} \rho_j(t) 
- \sum_{j=m+1}^n {S}_{ij} (\rho_j(t)+R_j^*) 
\right] \nonumber
\eea
where we have substituted the fixed point values of $R_i^*$ from
Eq.\ref{ooaacantona}.
When we drop the quadratic terms and change the independent
variable from $t$ to $\tilde{\alpha}$ by Eq.\ref{Bfns}, we obtain the
linearised system
\bea
\left[ \frac{d \rho_i(t)}{d \ln \tilde{\alpha}(t)} \right] _{i=1,\ldots ,m }
& \approx & -\frac{\rho_i(t)}{b} \left[
(r_i+b) - \sum_{j=m+1}^n S_{ij} R_j^* \right] \ \ \ \nn \\
\left[ \frac{d \rho_i(t)}{d \ln \tilde{\alpha}(t)} \right] _{i=m+1,\ldots,n }
& \approx & \frac{1}{b} R_i^*
\sum_{j=1}^n S_{ij} \rho_j(t)
\label{linearised1}
\eea
Eq.\ref{linearised1} then describes the behaviour of the trajectories as they
approach the fixed point. 

For the ${i=1,\cdots m}$ the equations are of the simple form:
\beq
\left[ \frac{d \rho_i(t)}{d \ln \tilde{\alpha}(t)} \right] _{i=1,\ldots ,m }
\approx \lambda_i \rho_i (t)
\eeq
where
\beq
\lambda_i=\frac{1}{b}\left[ \sum_{j=m+1}^n S_{ij} R_j^* - (r_i+b) \right]
\label{trivial}
\eeq
with solution,
\beq
[ \rho_i (t)]_{i=1,\cdots m} = a_i (\tilde{\alpha })^{\lambda_i }
\eeq
As in the previous argument, for $b>0$, in order to have
$\rho_i \rightarrow 0 $ in the infra-red we require $\lambda _i >0$,
where here $\lambda_i$ is given by Eq.\ref{trivial}.

For $i=m+1,\ldots,n $, assuming that the $m$ zero solutions are
infra-red stable, as discussed in the above paragraph,
we may simplify the procedure by working in the infra-red region where
$[\rho_i ]_{i=1,\cdots m} \rightarrow 0 $, so that we have
a simplified set of equations to solve:
\beq
\left[ \frac{d \rho_i(t)}{d \ln \tilde{\alpha}(t)} \right] _{i=m+1,\ldots,n }
 \approx  \frac{1}{b} R_i^*
\sum_{j=m+1}^n S_{ij} \rho_j(t)
\eeq
In fact the argument now parallels that given previously for the
non-zero solutions, as in Eq.\ref{linearised} and the discussion below
it, with attention now being focussed on the $(n-m)\times(n-m)$ 
lower right hand block of the re-ordered matrix $S$.

As a simple example of the application of these results,
we consider the example of the next-to-minimal supersymmetric
standard model (NMSSM) described by the superpotential:
\beq
W= h Qt^cH_2 + \lambda N H_1 H_2 -\frac{k}{3} N^3 
\eeq
where we have included in addition to the top quark Yukawa coupling
the NMSSM terms which contain the gauge singlet $N$.
The fixed points of this theory in the presence of the gauge couplings
have been previously studied numerically \cite{NMSSM}.
It was found there that the couplings $\lambda$ and $k$
are attracted towards zero values \cite{NMSSM},
and it is interesting to see how our analytic treatment here reproduces this.
The relevant RGEs are~\cite{ours,NMSSM}, keeping only the
QCD gauge coupling:
\bea
16\pi^2\frac{dk}{d\ln \mu} & = & k (6k^2 + 6 \lambda^2) 
\nonumber \\
16\pi^2\frac{d\lambda }{d\ln \mu} & = & \lambda (2k^2 + 4 \lambda^2 + 3h^2) 
\nonumber \\
16\pi^2\frac{dh }{d\ln \mu} & = & h(\lambda^2 + 6h^2 -2g^2(C_Q+C_{t^c}) )
\eea
In the notation of Eqs.\ref{convenient}, \ref{ratio} we find:
\bea
\frac{dR_k}{dt} & = & \tilde{\alpha}R_k [b -6R_k - 6 R_\lambda ] 
\nonumber \\
\frac{dR_\lambda }{dt} & = & 
\tilde{\alpha}R_\lambda [b -2R_k - 4 R_\lambda - 3R_h  ]
\nonumber \\
\frac{dR_h }{dt} & = & 
\tilde{\alpha}R_h[(r_h+b)   - R_\lambda - 6R_h ]
\eea
where $r_h=2(C_Q+C_{t^c})$.

We now order the ratios as $R_i=(R_k,R_\lambda ,R_h)$ and write the RGEs
for the ratios in the form of Eq.\ref{RGER} where we identify:
\beq
r_1=0, \ r_2=0, \ r_3= r_h = 16/3
\eeq
with $b=-3$ for the NMSSM as in the MSSM, and the matrix $S$ given by:
\beq
S = \left(
\begin{array}{ccc}
6 & 6  & 0 \\  
2 & 4  & 3 \\  
0 & 1  & 6 \\  
\end{array}        
\right)
\eeq
The non-zero couplings are obtained by inverting the matrix $S$,
then using Eq.\ref{nontrivial} with
$(r_j+b)=(-3,-3,7/3)$. The result is:
\beq
R_i^* = (\frac{29}{18},-\frac{19}{9},\frac{20}{27})
\eeq
The negative value of the fixed point for $R_\lambda$ means that we 
must abandon this fixed point as a physical possibility.

We next try a fixed point with $R_\lambda^* =0$, and the other two couplings
being given by their non-zero solutions. Following our procedure
we first re-order the couplings so that those with zero fixed point values are
listed first.
In this case we re-order the couplings as : $R_i=(R_\lambda , R_k ,R_h)$,
with $(r_j+b)=(-3,-3,7/3)$ (unchanged), but now we have:
\beq
S = \left(
\begin{array}{ccc}
4 & 2  & 3 \\  
6 & 6  & 0 \\  
1 & 0  & 6 \\  
\end{array}        
\right)
\eeq
Since by fiat $R_1^*=0$ it only remains to determine the 
fixed point values for the other two couplings. Since 
the lower $2 \times 2 $ block of the re-ordered matrix $S$,
is proportional to the unit matrix in this case it is easy to invert
and we find 
\beq
(R_k^*,R_h^* )= (-\frac{1}{2},\frac{7}{18})
\eeq
For $R_k^*=0$ with the others non-zero, we find
\beq
(R_\lambda^*,R_h^*)=(-\frac{25}{21},\frac{37}{63}).
\eeq
The trial fixed point in these cases again contains a non-physical negative
value, so we abandon these possibilities.

Finally we try a fixed point with $R_\lambda = R_k =0$, but with $R_h\neq 0$.
No further re-ordering is required since the two zero ratios occur for
$i=1,2$. The non-zero fixed point for the top Yukawa coupling
is given by:
\beq
R_h^*=R_3^*=S_{33}^{-1}(r_3+b) =\frac{7}{18}
\eeq
Having obtained a physically sensible fixed point
we must now test its infra-red stability. The behaviour of the couplings
$[R_i]_{i=1,2}$ around the origin is determined by 
the eigenvalues in Eq.\ref{trivial} which are given by:
\beq
[\lambda_i ]_{i=1,2} = \frac{1}{b} [ S_{i3}R_3^* - (r_i+b)] =
[-1,-\frac{25}{18}]
\eeq
which indicates that this fixed point is attractive in the infra-red
direction. 
The behaviour of the coupling $R_3$ around its fixed point is 
governed by the single entry matrix
\beq
A_{33}=\frac{1}{b}R_3^*S_{33} = -\frac{7}{9}
\eeq
whose eigenvalue is trivially equal to $\lambda = -\frac{7}{9}$.
Since $b<0$ in this case, a negative eigenvalue indicates an 
IRSFP\@.
In fact $R_h^*= 7/18$ is just the
usual Pendleton-Ross fixed point of the MSSM for the top quark 
Yukawa coupling discussed previously.

There are several important technical
questions not addressed by the above arguments.
One of these questions is: how fast~\cite{speed}
does the solution approach the fixed point?
So far, we can only suggest a numerical test
where the boundary conditions are picked from some range of initial parameters
spanning a range centred upon 1 and of the
order 1. We know that the IRSFP is only formally
realised in the
limit $\mu \rightarrow 0$.
While it is clear that when the magnitude of $\lambda_i$ is large, so is the
speed
of the running and therefore the rate at which the fixed point is
approached, in models of particle physics $\mu$
never reaches this limit and so the fixed point solution can only be used
approximately.
Some measure on the ``error'' of this prediction would then be
useful.
It is not clear yet if there are general classes of models which possess an  
IRSFP\@. Ideally, there would only be one IRSFP
in the theory. If two exist in the same model, it would be unclear a priori
which would be more likely to be realised in nature.
Finally, we should note that the solutions $R_{i>m}^*$ to Eq.\ref{ooaacantona}
are not all
guaranteed to be positive. Any that are negative will never be reached: the
coupling would simply approach zero and therefore should be added
to the list of couplings $R_{i=1,\ldots,m}$.
 
To summarise, we have determined the fixed points in a large class of
supersymmetric models. These fixed points offer the possibility of
increased predictivity and therefore testability of the models. There are
however there are several important theoretical questions raised by
this approach. Perhaps the most
important is that of the existence and uniqueness of IRSFPs in any given
theory. In the absence of such general theorems we have provided a well
defined procedure for exploring the fixed point properties. 
Our suggested procedure is the following: 
start with the case of a non-trivial
fixed point for all the $n$ physical couplings, 
and examine its stability; 
then go on to the case with one zero fixed point coupling and
$n-1$ non-zero fixed point couplings and examine its stability;
and so on down to the case of the completely trivial fixed point with $n$
zero fixed point couplings. 
We have the given conditions for infra-red stability 
for the general case of $m$ zero fixed point couplings
and $n-m$ non-zero fixed point couplings, which enables this procedure
to be followed.
Applying the analysis to the NMSSM in the low $\tan \beta$ regime, we are able
to show that the fixed point prediction for the top quark mass is equivalent
to the MSSM prediction. These techniques may also be applied
to the fermion mass problem \cite{us}.

\end{document}